\documentclass{aa}
\usepackage{txfonts}
\usepackage{graphicx}

\begin{document}

\title{The AGB stars of the intermediate-age LMC cluster NGC 1846}
\subtitle{Variability and age determination}

\author{T. Lebzelter\inst{1} \and P.R. Wood\inst{2}}

\institute{
Institute of Astronomy, University of Vienna, Tuerkenschanzstrasse 17, A1180 Vienna, Austria
\and
Research School for Astronomy \& Astrophysics, Australian National University, Weston Creek, ACT 2611, Australia}

\date{Received / Accepted}

\abstract{}{To investigate variability and to model the pulsational behaviour of AGB variables in the intermediate-age LMC cluster 
NGC 1846.}{Our own photometric monitoring has been combined with data from the MACHO archive to
detect 22 variables among the cluster's AGB stars and to derive pulsation periods. According to the
global parameters of the cluster we construct pulsation models taking into account the effect of
the C/O ratio on the atmospheric structure.  In particular, we have used opacities appropriate for
both O-rich stars and carbon stars in the pulsation calculations.}
{The observed P-L-diagram of NGC 1846 can be fitted using 
a mass of the AGB stars of about 1.8\,M$_{\sun}$. We show that the period of pulsation is
increased when an AGB star turns into a carbon star. Using the mass on the AGB defined by the pulsational
behaviour of our sample we derive a cluster age of $1.4\times10^{9}$ years. This is the first time
the age of a cluster has been derived from the variability of its AGB stars.  The carbon stars
are shown to be a mixture of fundamental and first overtone radial pulsators.}{}


\maketitle

\section{Introduction}

The cluster NGC 1846 belongs to an intermediate age population
in the Large Magellanic Cloud (LMC). In contrast to the Milky Way
the LMC harbours many clusters with an age around 1 to 5
Gyr (e.g. Girardi et al.\,\cite{GCBB95}). These clusters thus provide an interesting possibility
to study the late stages of stellar evolution for stars around 1.5
to 2.5 M$_{\sun}$, while the globular clusters belonging to the
Milky Way trace the evolution of stars of only up to about 0.9 M$_{\sun}$.

While there is a general agreement in the literature that NGC 1846 is
indeed an intermediate age cluster, the published values on global parameters 
of this system show some scatter. Derived metallicities can be roughly
divided into two groups, one around [Fe/H]$=$-1.5 (Dottori et al.\,\cite{DPB83},
Leonardi \& Rose \cite{LR03}) and one around [Fe/H]$=$-0.7 (Bessell et al.\,\cite{BWL83}, 
Bica et al.\,\cite{BDP86}, Olszewski et al.\,\cite{OSSH91}, Beasley et al.\,\cite{BHS02}). 
 Probably the first determination of the metallicity was done by
Cohen (\cite{C82}) giving a value in between the two discussed metallicity levels, namely [Fe/H]$=$-1.1.
The most recent determination of the cluster's metallicity is given by  Grocholski et al.\,(\cite{GCS06}), 
who derive a value of [Fe/H]$=-$0.49$\pm0$ based on the Ca triplet strength of 17 individual cluster
members measured with the VLT.

Age values for NGC 1846 scatter between less than 1 Gyr (Frantsman \cite{F88})
and 4.3 Gyr (Bica et al.\,\cite{BDP86}). Mackey \& Broby Nielsen (\cite{MB07}) give
cluster ages from 1.5 to $2.5 \times 10^{9}$ years based on a comparison of the colour
magnitude diagram and theoretical isochrones. The range in age thereby results from the usage of
two different sets of isochrones. These ages correspond to masses on
the AGB of about 1.3 to 1.8 M$_{\odot}$. Mackey \& Broby Nielsen (\cite{MB07}) also
report on the probable existence of two populations in NGC 1846 with similar metallicity
but separated in age by about 300 Myr.

A reddening of $A_{V}$$=$0.45\,mag has been determined by Goudfrooij et al.\,(\cite{GGKM06}).
Grocholski et al.\,(\cite{GCS06}) note that NGC 1846 is probably suffering from differential
reddening without giving any detailed numbers. Keller \& Wood (\cite{KW06}) give a somewhat
lower reddening of $E(B-V)$$=$0.08, i.e. an $A_{V}$ value of 0.25\,mag.

The first studies of the stellar content of NGC 1846 were published by Hodge (\cite{H60}) and Hesser
et al.\,(\cite{HHU76}). A number of luminous stars on the Asymptotic Giant Branch (AGB) were 
identified and published with finding charts by Lloyd-Evans (\cite{LE80}). Throughout this paper we
will use the naming given in Lloyd-Evans' publication (LE{\it XX}) except for H39, which 
follows the numbering from Hodge (\cite{H60}). Details on individual AGB stars have been 
published in a number of papers. Frogel et al.
(\cite{FPC80}, \cite{FMB90}) and Aaronson \& Mould (\cite{AM85}) gave near infrared photometry and 
$m_{\rm bol}$ values for most of Lloyd-Evans' AGB stars and a few more cluster objects, 
 unfortunately without any finding chart. Tanab\'{e} et al.\,(\cite{T98}) searched this cluster
for extreme infrared stars, i.e.~stars with a high circumstellar absorption in the visual range,
but found none.

No investigations on the variability of the AGB stars in NGC 1846 has
been published up to now. However, as most AGB stars are pulsating (forming the group of
long period variables or LPVs) it seemed very likely that
some light variability would be found in these stars. 
In a previous paper (Lebzelter \& Wood \cite{LW05}) 
we showed that the variability analysis of LPVs in a single stellar 
population like the globular cluster 47 Tuc allows one to investigate various fundamental aspects of AGB stars 
like the evolution of the pulsation mode or mass loss. With NGC 1846 we have chosen an
interesting alternative target since its AGB stars are more massive and include a number of carbon stars.

\section{Observations and data reduction}

For the detection of AGB variables in NGC 1846 and the determination of their pulsation period
two data sets were used. We started a photometric monitoring program using ANDICAM
at the 1.3m telescope on CTIO which belongs to the SMARTS 
consortium\footnote{www.astronomy.ohio-state.edu/ANDICAM/}. The plan was to
obtain images of the cluster in $V$ and $I$ at a rate of approximately twice a week. However,
this plan was not feasible due to a strong pressure on the observing time at this instrument. 
Still we got a total of 63 epochs between JD 2453584 and JD 2454048 with a gap between JD 2453856
and JD 2453950.

We used the pipeline reduced $V$ and $I$ images provided by the SMARTS consortium. 
The detection of variables and the extraction of the light curves was done using the image 
subtraction tool ISIS 2.1 (Alard \cite{Alard00}). The resulting light curves were searched for periodicities using
Period98 (Sperl \cite{Sperl98}). A maximum of three periods was used to fit the light
change. 

It turned out that the data set obtained at CTIO was very valuable to detect periods below 100
days, but that it was too short for
a reliable determination of the long periods occurring in some of the AGB stars of NGC 1846.
To improve our period determination we thus decided to access the MACHO database as
the cluster was covered by this survey.  We used the coordinates of the variable AGB stars
detected within our CTIO time series to extract the corresponding MACHO light curves.
Two additional long period variables not apparent in the CTIO data were also found.
The MACHO data consisted of about 250 epochs over the time span between JD 2449001 and JD 2451546,
i.e. about 2500 days. The cluster was in the overlap region of two MACHO frames so that for
part of the sample two series of measurements were available. 
Periods were determined in the same way as for the CTIO data. Data both from the blue and
from the red pass band were analyzed separately and gave very similar results. Also, time series from the
two different MACHO frames resulted in the same periods within a few percent.
Due to the significant separation in time and due to 
the different filter systems no attempt was made to combine the MACHO and the CTIO data.

Periods determined from both data sets were then compared with each other. The agreement for
periods shorter than about 100 days was quite good (differing by a maximum of 10 percent) 
taking into account the significant irregularities found on top of the regular light change. An average period
was calculated from the two data sets and used in the further analysis. For LE16 and a previously
unidentified variable, LW1, the MACHO light curve was not useable. For H39 no counterpart could
be identified in the MACHO data base. All longer periods
were taken from the MACHO data. Compatibility of the MACHO period with the data set from
CTIO was checked by eye in all these cases.

The uncertainties of the periods, calculated with a least square fit approach, are less than a few percent.
For the MACHO data set, photometric errors were taken from the MACHO database. Typical values are
below 0$\fm$01. For the CTIO data set we estimate a typical error of 0$\fm$01.

As part of the CTIO time series we obtained near infrared time series in $J$ and $K$
using the same instrument. This data set covers roughly three months and allows us to estimate the light
variability in the near infrared. Standard data reduction was applied to the near infrared images using IRAF.
The background was subtracted using dithering.
The $J$ and $K$ magnitudes were transformed to the 2MASS system using the transformation from
Carpenter (\cite{Carpenter01}). A set of reference stars within the cluster with near infrared photometry data
from 2MASS was chosen to set the zero point for each epoch. These stars remained constant within $\pm$0$\fm$02
relative to each other on all frames. This value can also be seen as the typical accuracy achieved for the
variables. 

Additional $J$ and $K$ photometry was obtained with the near-infrared array camera CASPIR
(McGregor et al. \cite{MHHB94}) on the 2.3m telescope of the Australian
National University at Siding Spring Observatory.  The $J$ and $K$ magnitudes on the
CASPIR system were converted to the 2MASS system using the transformations in
McGregor (\cite{McGregor94}) and Carpenter (\cite{Carpenter01}). From these data we also
obtained near infrared photometry for a large number of non variable, low luminosity AGB
and RGB stars required for calibrating the models (see below).

\begin{figure}
\resizebox{\hsize}{!}{\includegraphics{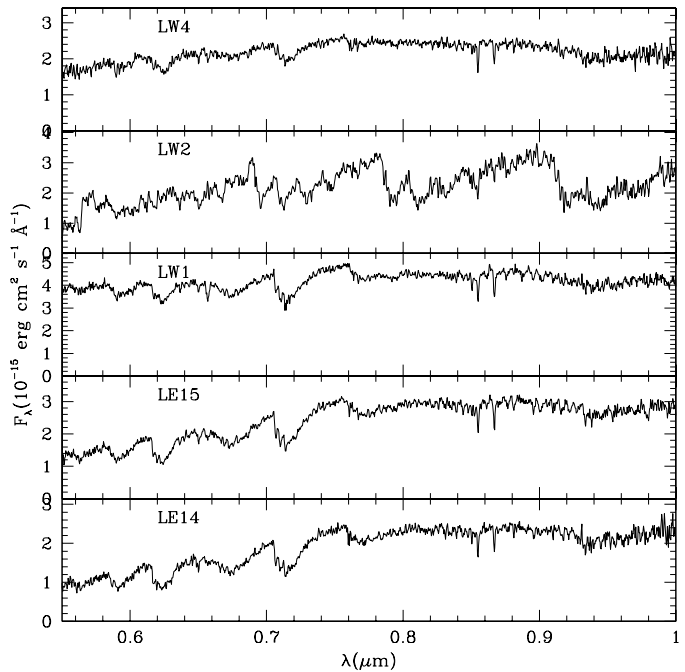}}
\caption{Spectra of the NGC 1846 AGB variable stars without a previous
spectral type.}
\label{newspecs}
\end{figure}

For 5 variable stars with a previously unknown spectral type, spectra were
obtained with the Dual Beam Spectrograph on the ANU 2.3m telescope at Siding
Spring Observatory.  The spectra are shown in Fig.\,\ref{newspecs} and the spectral types
are listed in Table \ref{sample}.  
For the O-rich stars, the spectral types were based on the strengths of the TiO
bands at 616\,nm and 705\,nm.  LW4 appears to be of late K spectral type while LW1, LE14
and LE15 are of early M spectral type.  The C$_{2}$ bandhead at 564\,nm and the strong
CN bandheads at 692 and 788\,nm clearly identify LW2 as a carbon star (see Turnshek et al.
\cite{Turnshek85} for an atlas of K, M, S and C star spectra).

\section{Results}
A total of 23 red variables were detected within the CTIO/ANDICAM field. Nineteen of these
could be identified with AGB stars found by Lloyd-Evans (\cite{LE80}). For two of the stars, LE9 and LE18
variability was detected, but no period could be determined. Four further variables had no previous identifier in the literature.
Their coordinates are given in Tab.\,\ref{newcoords}.
Similar to our approach in Lebzelter \& Wood (\cite{LW05}), we named them with a prefix LW followed by a number.

\begin{figure*}
\sidecaption
\includegraphics[width=12cm]{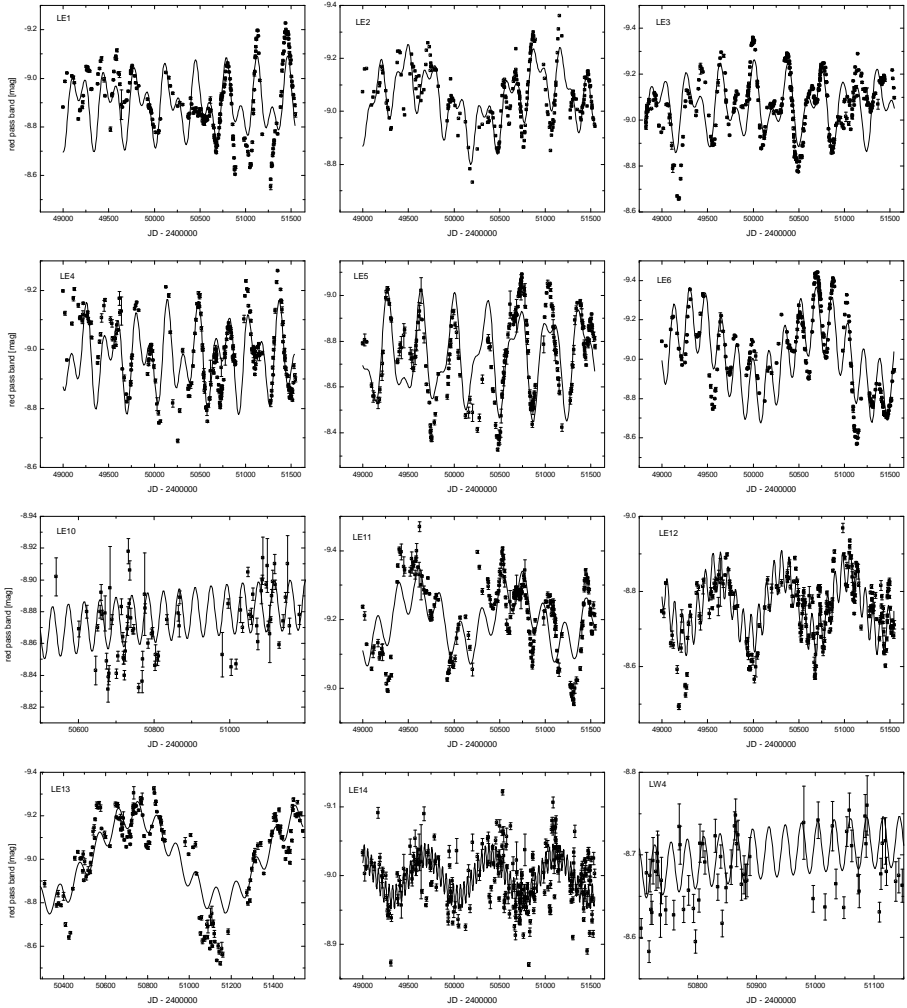}
\caption{MACHO light curves of the variables of our survey showing long periods. Fourier fits with up to
two periods are shown. Note that the y-axis scale varies from star to star. For LE13 only part of the 
light curve is plotted for a better illustration of the short period light change.}
\label{lcmacho}
\end{figure*}

\begin{table}
\centering
\caption{Coordinates from the 2MASS catalogue for variables with no identifier 
in the literature. For cross correlation with infrared sources
identified by Tanab\'{e} et al.\,(\cite{T04}) see sect.\,\ref{discus}).}
\begin{tabular}{lccl}
\hline
Ident & RA (2000) & DE (2000) & comment\\
\hline
LW1 & 05:07:36.96 & -67:27:56.9 & \\
LW2 & 05:07:40.26 & -67:27:40.3 & \\
LW3 & 05:07:54.46 & -67:24:46.2 & prob.~non-member\\
LW4 & 05:07:34.15 & -67:27:48.5 & \\
\hline
\noalign{\smallskip}
\end{tabular}
\label{newcoords}
\end{table}

\begin{figure*}
\sidecaption
\includegraphics[width=12cm]{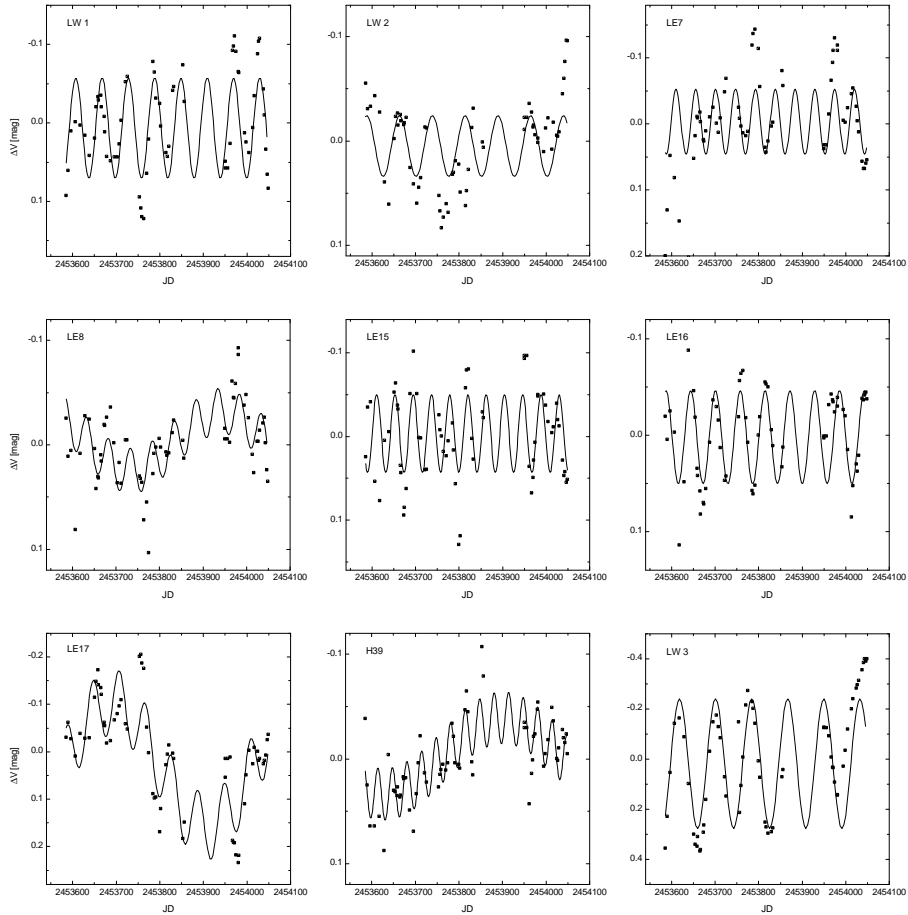}
\caption{ANDICAM/CTIO light curves of the variables of our survey dominated by short periods. Fourier fits with up to
two periods are shown. Note that the y-axis scale varies from star to star. LW3 is probably
not a member of NGC 1846 (see text).}
\label{lcandi}
\end{figure*}

For stars with their light change
dominated by long periods, MACHO light curves with corresponding Fourier fits are presented in Fig.\,\ref{lcmacho}.
For the shorter period stars we show the CTIO data with the corresponding Fourier fits in
Fig.\,\ref{lcandi}. LW3, shown in Fig.\,\ref{lcandi}, is obviously a nicely periodic (83 d), red ($(J-K)_{2MASS}=$1.17)
variable but, as it is located more than 3.5 arcmin away from the cluster center, we have to strongly doubt its
membership to the cluster. 
%Similarly, LMC-BM 13-12, a known C-star, is probably not a cluster member.
%We show no light curve for this star here, but is clearly variable as suspected already earlier by Wright \& Hodge
%(\cite{WH71}). 
This star will be excluded from the further analysis. Thus we have now a sample of 22 red variables associated with 
NGC 1846, 20 of them showing a periodic light change.
The determined periods of these stars together with other basic data are listed in Tab.\,\ref{sample}.

At a first glance two features of our sample are remarkable. First, multiperiodicity seems
to be a common phenomenon among these stars. Only the weakest stars of our sample could be fitted with
a single period. Thereby we observe two kinds of multiple periods, on the
one hand two periods with a ratio of about 2, and on the other hand 
long secondary periods, a well known feature of AGB variables in the Galactic field and in
the MCs (e.g. Wood et al.\,\cite{wok04}, Wood\,\cite{wood07}). This long secondary period can be the dominant one 
as can be seen e.g. in LE12 and LE13 (see Fig.\,\ref{lcmacho}). Second, there are no
large amplitude stars in NGC 1846, i.e.~the classical miras seem to be missing. The light variations
in the visual hardly exceed 1 mag. However, we find a considerable fraction of stars in the period
range typically occupied by the miras (i.e. around 300 days).

\begin{figure}
\centering
\resizebox{\hsize}{!}{\includegraphics*{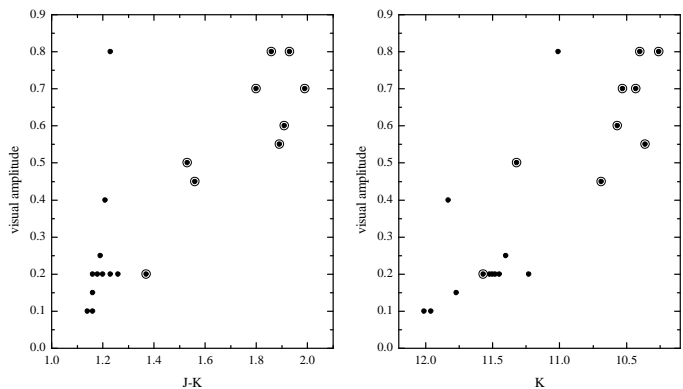}}
\caption{Amplitude vs.~$K$ (right panel) and $J-K$ (left panel). No correction has been obtained for the difference
in the filter passband between the MACHO and the ANDICAM time series. The amplitude given corresponds to the maximum
light difference found in our time series. O-rich and C-rich stars are marked by dots and circled dots, respectively.}
\label{tanabe}
\end{figure}

Tanab\'{e} et al.~(\cite{tana04}) found a rapid increase in light amplitude once the star reaches the tip
of the AGB. We see a similar relation between the light amplitude and mean $K$ and $J-K$ values (Fig.\,\ref{tanabe}).

\begin{figure}
\centering
\resizebox{\hsize}{!}{\includegraphics*{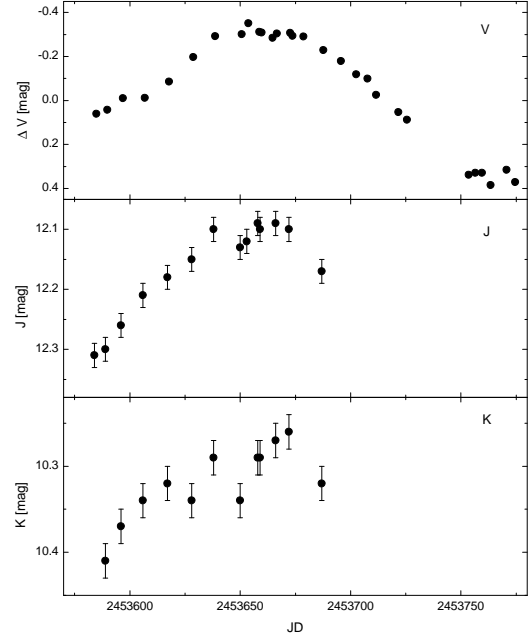}}
\caption{Near infrared light variations of LE5 in $J$ (middle panel) and $K$ (lower panel) from CTIO/ANDICAM data.
For comparison the visual light curve (CTIO/ANDICAM) for the same time span is shown (upper panel).}
\label{lcir}
\end{figure}

The near infrared measurements show only a slight scatter typically of the order of the
estimated error of the individual measurements. This is not surprising given
the low light amplitudes found in the visual data. One star, LE 5, clearly exhibits
a light change above the error level. Its light curve is shown in Fig.\,\ref{lcir}. 
The same behaviour of the light change as in the visual can be seen clearly. 
As the time series spans only a time interval of
about 3 months long time changes are not properly covered. Thus we cannot exclude that the
true light amplitude in the near infrared is larger than our estimate.
The
variation of about 0.2 mag seen for LE5 in $J$ and about 0.15 mag in $K$ correspond to about 0.4 mag
in the visual light curve. As the light amplitude of our sample stars hardly exceeds 1 mag in
the visual we can estimate that the total amplitude in $J$ and $K$ will be less than 0.6
and 0.4 mag, respectively. 

Near infrared colour changes are also very small (see Fig.\,\ref{lcir}). We used our time series to calculate
mean $K$ and $J-K$ values for the further analysis.  Our values are in
reasonable agreement with the 2MASS data.

\begin{table}
\caption{The AGB variables of NGC 1846. Reddening corrected photometry values are given. Periods are given in
days. Multiple periods are listed in consecutive lines. A colon
marks an uncertain period. Spectral types (last column) are from Frogel et al.\,(\cite{FMB90}, see references
therein) except for LE8 and LE13 which are taken from Lloyd-Evans\,(\cite{LE83}). LW3 is probably not a
cluster member and thus not listed.}
\begin{tabular}{lrccccc}
\hline
Ident & $P$ & $K$ & $J-K$ & log\,$L$/L$_{\odot}$ & log\,$T_{\rm eff}$ & Spec.\\
\hline
LE1 & 170 & 10.53 & 1.80 & 3.859 & 3.428 & C\\
 & 325 & & & & & \\
LE2 & 150 & 10.57 & 1.91 & 3.835 & 3.411 & C\\
 & 289 & & & & & \\
LE3 & 189 & 10.43 & 1.99 & 3.888 & 3.397 & C\\
 & 351 & & & & & \\
LE4 & 173 & 10.36 & 1.89 & 3.920 & 3.414 & C\\
 & 307 & & & & & \\
LE5 & 188 & 10.26 & 1.93 & 3.958 & 3.407 & C\\
 & 355 & & & & & \\
LE6 & 172 & 10.40 & 1.86 & 3.905 & 3.417 & C\\
 & 1401 & & & & & \\
LE7 & 47 & 11.40 & 1.19 & 3.670 & 3.565 & M\\
 & 650 & & & & & \\
LE8 & 51 & 11.23 & 1.23 & 3.719 & 3.557 & S\\
 & long & & & & & \\
LE10 & 41 & 11.96 & 1.16 & 3.463 & 3.571 & K\\
LE11 & 227 & 10.69 & 1.56 & 3.842 & 3.470 & C\\
 & 1066 & & & & & \\
LE12 & 86 & 11.32 & 1.53 & 3.598 & 3.476 & C\\
 & 716 & & & & & \\
LE13 & 92 & 11.01 & 1.23 & 3.806 & 3.557 & S\\
 & 837 & & & & & \\
LE14 & 60 & 11.52 & 1.20 & 3.618 & 3.563 & M\\
 & 715: & & & & & \\
LE15 & 41 & 11.48 & 1.26 & 3.607 & 3.553 & M\\
 & 32: & & & & & \\
LE16 & 57 & 11.50 & 1.16 & 3.645 & 3.570 & M\\
LE17 & 61 & 11.83 & 1.21 & 3.491 & 3.562 & M\\
H39 & 33 & 11.77 & 1.16 & 3.538 & 3.571 & M\\
LW1 & 60 & 11.45 & 1.18 & 3.655 & 3.567 & M\\
LW2 & 75 & 11.57 & 1.37 & 3.545 & 3.506 & C\\
LW4 & 27 & 12.01 & 1.14 & 3.448 & 3.573 & K\\
\hline
\noalign{\smallskip}
\end{tabular}
\label{sample}
\end{table}

\section{Pulsation models}
Given the large number of LPVs now known in NGC 1846, we can now attempt to
model the period-luminosity (PL) laws.  For our models, we adopt the following
parameters: distance modulus $(m-M)_{V}$=18.54 and reddening $E(B-V)$$=$0.08
(Keller \& Wood \cite{KW06}), helium mass fraction Y=0.27 and a metal abundance
Z=0.004 (following Grocholski et al. \cite{GCS06}).   

Below, we derive the mass of the AGB stars by fitting their period-luminosity
(PL) relations, using the values mentioned above as a guide only.  We note
that the variables in this cluster are mostly of relatively small amplitude so
that high mass loss rates are unlikely to have begun.  It is also known that
red giant mass loss prior to the superwind phase in intermediate mass stars
with $M \sim 1.5$\,M$_{\odot}$ is much less significant than in lower mass red
giant stars such as those in globular clusters.  We therefore assume a constant
mass along the AGB where our variables are found.

The static and linear non-adiabatic pulsation models were created with updated
versions of the pulsation codes described in Fox \& Wood (\cite{fw82}).  For
temperatures above $\log T = 3.75$, the interior opacities of Iglesias \& Rogers
(\cite{IR96}) were used, including tables with the carbon enhancements required for
carbon stars.  For temperatures below $\log T = 3.75$, a scheme similar to that
of Marigo (\cite{Marigo02}) was used for both oxygen-rich and carbon-rich opacities.
Molecular contribution due to CN, CO and H$_{2}$O and a contribution due to
metals of low ionization potential were added to low temperature opacities for
zero-metallicity mixtures of Alexander \& Ferguson (\cite{AF94}).  Details will be
given elsewhere.  The core mass $M_{\rm c}$ was obtained from the $L$-$M_{\rm
c}$ relation of Wood \& Zarro (\cite{WZ81}).

\begin{figure}
\centering
\resizebox{\hsize}{!}{\includegraphics{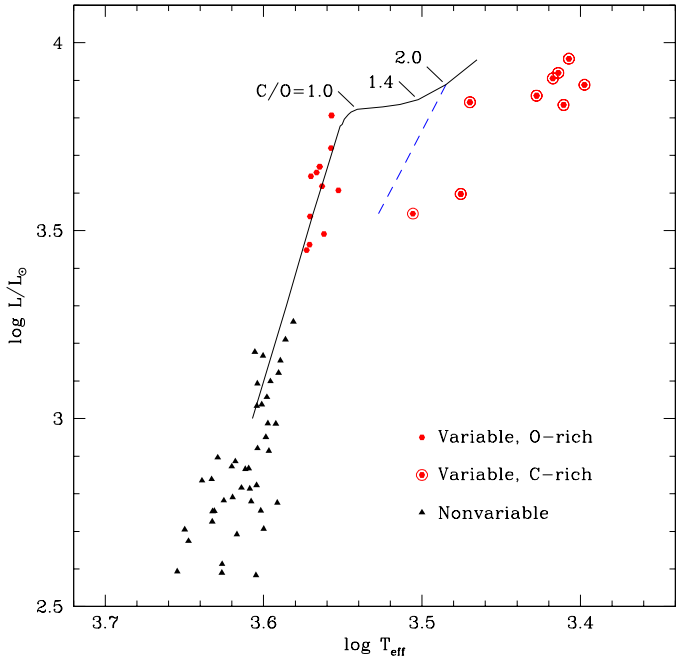}}
\caption{The tip of the NGC\,1846 giant branch in the HR-diagram.  The variable
and non-variable stars are shown along with their spectral types.  The solid line
is the theoretical giant branch, which has C/O increasing when $L$ is
greater than 6000 L$_{\odot}$ (see text).
Values of C/O are indicated at several points on the theoretical track.
The dashed line is the line which a C-rich star will follow when its
luminosity declines and rises during a helium shell flash cycle. }
\label{hrd}
\end{figure}

Since we are computing pulsation periods, it is important to get the radii (and
hence $T_{\rm eff}$) of the models correct.  In order to produce the correct
$T_{\rm eff}$ on the AGB, the model AGB had to be made to coincide with the AGB
of NGC\,1846.  The observational AGB was obtained by converting $K$ and $J$-$K$
of the variables and non-variables in the cluster to $M_{\rm K}$ and
($J$-$K$)$_{0}$ using the distance modulus and reddening given above.  Then,
for O-rich stars, $M_{\rm K}$ and ($J$-$K$)$_{0}$ were converted to
$\log L$/L$_{\odot}$ and $\log T_{\rm eff}$ using the transforms in Houdashelt
et al. (\cite{HBS00a}; \cite{HBS00b}) (the 2MASS J and K values were first transformed to the
SAAO system, similar to that of Houdashelt et al., using the transforms in
Carpenter \cite{Carpenter01}).  For C-rich stars, $M_{\rm K}$ and ($J$-$K$)$_{0}$ were
converted to $\log L$/L$_{\odot}$ and $\log T_{\rm eff}$ using the
\mbox{(($J$-$K$)$_{0}$, $T_{\rm eff}$)} relation from Bessell, Wood \& Lloyd Evans
(\cite{BWL83}) and the bolometric correction to $K$ from Wood, Bessell \& Fox (\cite{WBF83}).
Figure~\ref{hrd} shows the NGC\,1846 stars in the theoretical HR-diagram.

In order to match the models to the observed AGB, the mixing-length in the
convection theory was set so that the O-rich, Z=0.004 models matched the
$T_{\rm eff}$ of the O-rich stars in NGC\,1846. As noted above, all our models
have the same mass of 1.8\,M$_{\odot}$: as explained below, this mass is required 
in order to get the correct
pulsation periods.  It was found that a constant
mixing-length gave an AGB that decreased its $T_{\rm eff}$ too rapidly with
increasing $L$ (this is a general problem with all models e.g. the tracks or
isochrones of Girardi et al. \cite{GBBC00}).  It was found that the ratio of
mixing-length to pressure scale height $\alpha$ needed to increase slowly with
$L$ in order to obtain the correct slope for the theoretical giant branch
($\alpha$ was found to vary linearly with $L^{0.3}$).  This can be seen in
Table~\ref{puls_mods}.

It is clear from Figure~\ref{hrd} that above about $\log L/$L$_{\odot} = 3.8$
($L = 6310 $L$_{\odot}$), all stars in NGC\,1846 are C stars.  For $L > 6000
$L$_{\odot}$, we therefore assumed that C/O increased linearly with $\log
L/$L$_{\odot}$: this corresponds to a constant rate of carbon dredge up since
$\frac{d \log L}{dt}$ averaged over a shell flash cycle is constant for AGB
stars (Wood \cite{Wood74}).  We assumed that C/O reached 3 at $L = 9000 $L$_{\odot}$,
roughly the luminosity of the brightest AGB star in the cluster.  We have no
estimates of C/O for the stars in the cluster but values of C/O $\sim$ 3 at the
end of AGB evolution are indicated by planetary nebulae in the LMC
(Stanghellini et al.\,\cite{SSG05}).  

The model AGB created with the varying C/O ratio and C-rich opacities is shown
in Figure~\ref{hrd}.  As soon as C/O passed through 1.0, $T_{\rm eff}$
decreased rapidly as the CN molecule and its opacity became more prominent.
When C/O $\sim$ 1.3, essentially all the N atoms are bound to C atoms which are
not bound up in CO.  Thus the decrease in $T_{\rm eff}$ is not as rapid with
further increases in C/O.

It is notable that the model $T_{\rm eff}$ values are not as cool as the
$T_{\rm eff}$ values estimated for the observed stars by converting $J$-$K$ to
$T_{\rm eff}$.  As shown below, the model $T_{\rm eff}$ values lead to quite a
good match between observed and predicted pulsation periods, suggesting that
the model $T_{\rm eff}$ values are reliable.  This would indicate that the most
luminous and cool C stars in NGC\,1846 are warmer than estimated here.  This
could be because they have significant circumstellar shells leading to
reddening of the stars, or because the conversion from $J$-$K$ to $T_{\rm eff}$
is incorrect.

The dashed line in Figure~\ref{hrd} is the track that would be followed by a
C-rich star (with C/O = 2 in the case shown) undergoing a helium shell flash.
During the He-burning part of the shell flash cycle, AGB stars of mass 1-2
M$_{\odot}$ decline in luminosity by roughly a factor of 2 (Vassiliadis \&
Wood \cite{VW93}) and spend a reasonable fraction of the cycle time near the low
luminosity part of the cycle.  We suggest that the two stars with $\log
(L$/L$_{\odot}) < 3.7$ and $\log T_{\rm eff} < 3.52$ are stars in this phase
(both stars are C stars).

\begin{table*}
\caption{The pulsation models}
\begin{tabular}{rcccccrrrr}
\hline
$L$/L$_{\odot}$ & $M$/M$_{\odot}$ & $M_{\rm c}$/M$_{\odot}$ & $\ell$/H$_p$ & log\,$T_{\rm eff}$ & C/O & P$_0$ & P$_1$ & P$_2$ & P$_3$ \\
\hline
\multicolumn{10}{l} {AGB stars with $L$ near the H-burning maximum of the flash cycle} \\
  6000 &  1.8 &  0.596 &  1.882 &  3.5516 &  0.324 &   167.7 &     88.7 &    56.8 &    49.1 \\
  6069 &  1.8 &  0.597 &  1.884 &  3.5503 &  0.400 &   171.6 &     90.4 &    57.9 &    50.2 \\
  6256 &  1.8 &  0.601 &  1.890 &  3.5490 &  0.600 &   179.1 &     93.5 &    59.8 &    52.2 \\
  6448 &  1.8 &  0.604 &  1.895 &  3.5465 &  0.800 &   188.8 &     97.5 &    62.4 &    54.8 \\
  6547 &  1.8 &  0.605 &  1.897 &  3.5447 &  0.900 &   195.0 &    100.0 &    64.1 &    56.4 \\
  6647 &  1.8 &  0.607 &  1.900 &  3.5409 &  1.000 &   204.6 &    104.0 &    66.9 &    59.5 \\
  6748 &  1.8 &  0.609 &  1.903 &  3.5248 &  1.100 &   235.4 &    117.1 &    77.9 &    70.0 \\
  6851 &  1.8 &  0.611 &  1.905 &  3.5139 &  1.200 &   260.4 &    127.1 &    86.8 &    76.8 \\
  7062 &  1.8 &  0.614 &  1.911 &  3.5025 &  1.400 &   293.3 &    139.4 &    98.8 &    85.2 \\
  7734 &  1.8 &  0.626 &  1.927 &  3.4849 &  2.000 &   371.4 &    166.1 &   126.8 &   104.0 \\
  9000 &  1.8 &  0.647 &  1.955 &  3.4652 &  3.000 &   523.7 &    210.8 &   175.7 &   140.1 \\
\multicolumn{10}{l} {AGB stars in the He-burning phase of the flash cycle} \\
  7734 &  1.8 &  0.626 &  1.927 &  3.4849 &  2.000 &   371.4 &    166.1 &   126.8 &   104.0 \\
  4500 &  1.8 &  0.626 &  1.836 &  3.5140 &  2.000 &   163.3 &     89.3 &    59.1 &    53.4 \\
  3500 &  1.8 &  0.626 &  1.800 &  3.5276 &  2.000 &   114.9 &     66.4 &    43.6 &    31.5 \\
\hline
\noalign{\smallskip}
\end{tabular}
\begin{flushleft}
Notes:  All models helium abundance Y = 0.26.  The initial metal abundance (before third dredge-up)
is Z = 0.004.  $\ell$/H$_p$ is the ratio of mixing-length to pressure scale height.  P$_0$,  P$_1$, P$_2$ and P$_3$
are the linear nonadiabatic periods (in days) of the fundamental, 1st, 2nd and 3rd overtone modes.
\end{flushleft}
\label{puls_mods}
\end{table*}
 
\begin{figure}
\centering
\resizebox{\hsize}{!}{\includegraphics{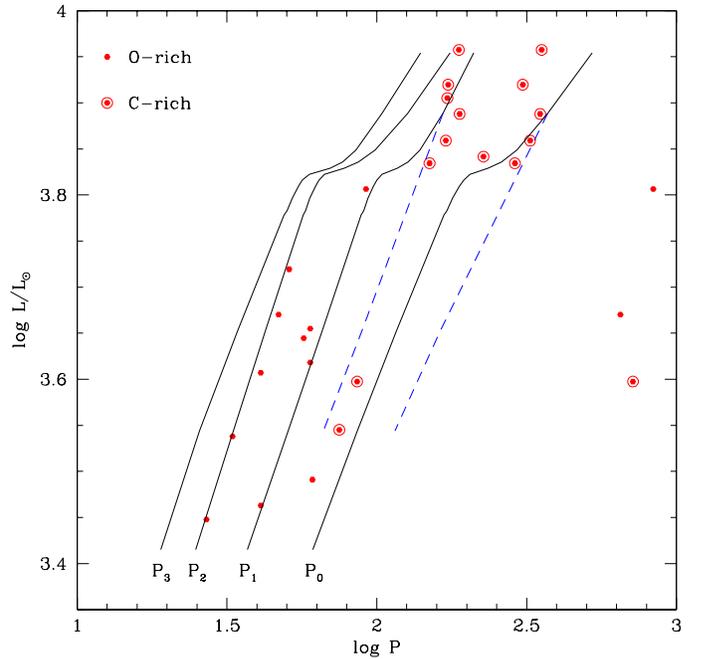}}
\caption{The $\log $($L$/L$_{\odot}$)-$\log P$ diagram for NGC\,1846 stars and models. The fundamental
mode and the first three overtones are shown for models with $L$ near the 
H-burning maximum of the He shell flash cycle (solid lines).  For AGB stars in the He-burning phase of the flash cycle,
only the fundamental and first overtone modes are shown (dashed lines).}
\label{pl_fig}
\end{figure}

\section{Discussion} \label{discus}
Having matched the O-rich models to the observed values of $T_{\rm eff}$, we
ensure that the models matched the observed pulsation periods of the O-rich
stars.  If not, the mass was altered, and new calculations of the giant branch
track and the pulsation periods were made.  It was found that a mass of
1.8\,M$_{\odot}$ gave the best fit between observations and theory for the
O-rich stars.  This corresponds to a cluster age of $1.4\times10^{9}$ years
using the isochrones of Girardi et al. (\cite{GBBC00})\footnote{Using the BaSTI
isochrones (Bedin et al. \cite{basti05}) we find an age of $1.9\times10^{9}$ years for a 1.8\,M$_{\odot}$
star to reach the AGB (using the same parameters as Mackey \& Broby Nielsen \cite{MB07}).}.  
The fit is shown in
Figure~\ref{pl_fig} and the model details are given in Table~\ref{puls_mods}.
It is clear that all the O-rich stars, with the exception of LE\,17 near $\log
P = 1.8$ and $\log L$/L$_{\odot} = 3.49$, are first or second overtone
pulsators (the stars with long secondary periods are not considered here).

It is clear from the models that as the C/O ratio increases beyond 1.0, the
periods of all modes increase rather rapidly due to the increase in radius and
decrease in $T_{\rm eff}$ caused by the molecules in the C-rich atmospheres.
Furthermore, it can be seen that the C stars fall quite nicely on two sequences
corresponding to pulsation in the first overtone and fundamental modes.  In the
absence of these C-rich models, it would have been natural to explain the C
stars as fundamental mode pulsators scattering broadly around an extension of
the fundamental mode O-rich sequence.

The fit of the models to the observed C star points is not perfect.  This is
not surprising given the rather elementary method used to compute the C-rich
opacities, combined with the fact that we do not really know the C/O ratios in
these stars.  Furthermore, as mentioned above, the stars move up and down in luminosity during a
He-shell flash cycle by factors of about 2 in luminosity (Vassiliadis \& Wood
\cite{VW93}).  The dashed lines in Figure~\ref{pl_fig} show how the fundamental and
first overtone periods would vary over a shell flash cycle for a star with C/O
= 2.  It seems most likely that the two C stars with $\log L$/L$_{\odot} < 3.6$
are near the luminosity minimum of a shell flash cycle (where AGB stars spend
about 20\% of their time).

In general, the theoretical and observed periods for the C-stars agree
reasonably well.  This indicates that the radii and $T_{\rm eff}$ values of the
models agree well with the real values for these stars and the assumption that
the temperatures indicated by the near infrared colours are misleading.  
%On the other hand, the
%$T_{\rm eff}$ values of the models are much hotter than the $T_{\rm eff}$
%values deduced for the C stars from their $J$-$K$ photometry.  This suggests
%that the C stars have a very significant circumstellar reddening, or that the
%($J-K$,$T_{\rm eff}$) conversion is invalid.

It is somewhat strange that the most luminous of the C stars appear to have
shorter pulsation periods than the less luminous C stars.  One way for this to
occur would be to decrease the C/O ratio in the most luminous C stars by
hot-bottom burning.  However, at 1.8 M$_{\odot}$, the mass is too low for
hot-bottom burning, since AGB masses greater than about 4
M$_{\odot}$ are required for hot-bottom burning to be significant.  Another
explanation may be that the nonlinear pulsation period differs from the linear
pulsation period when the pulsation amplitude increases, as noted in Lebzelter
\& Wood (\cite{LW05}). The visual light amplitudes give no indication for this
as the amplitudes of the carbon stars are all very similar (see Fig.\,\ref{tanabe}). 
In the infrared,
as shown in Fig.\,\ref{lcir}, the star with the largest amplitude in our data
set is LE5 which is also the most luminous star of our sample. This may
indeed indicate that the shorter periods at the tip of the AGB can be explained
by nonlinear effects. However, as mentioned above, our near infrared time series
are too short for a definite conclusion on this point.

A reduction in stellar mass due to significant amounts of mass loss can not
explain the shorter periods of the most luminous C stars since this would {\em
increase} the pulsation period.  In general, we have not found it necessary to
invoke any significant reduction in mass with luminosity up the AGB in these
calculations.  
We note that this contrasts with the stars in 47\,Tuc studied by Lebzelter  
\& Wood (\cite{LW05}) where a significant amount of mass loss was found on the
AGB.  As noted briefly in Section 4, this is as expected.  The stars being
studied here and in 47\,Tuc appear to have relatively low mass loss rates where
a Reimers' mass loss law may apply.  In such a case, the mass loss rate is
proportional to $LR/M$.  For a given $L$, this functional dependence suggests that 
the intermediate mass ($\sim$1.8 M$_{\odot}$)
stars in NGC\,1846 will have lower mass loss rates than the low mass ($\sim$0.9 
M$_{\odot}$) 47\,Tuc stars because of the higher mass.  In addition, the
giant branch temperature increases with stellar mass so the radius will be smaller
at a given $L$ in NGC\,1846.
The consequence of the expected lower mass loss rate in NGC\,1846 is that a smaller 
amount of mass is lost on the giant branches up to a given $L$.  The fractional
change in mass in NGC\,1846 is even smaller because of the higher mass.

%Once the superwind mass loss phase begins, which appears to lie beyond the C star phase
%observed here in NGC\,1846, the
%AGB mass will be significantly reduced and the period will increase significantly
%because of this (since $P \propto M^{-0.8}$ to a good approximation for the
%fundamental mode e.g. Fox \& Wood \cite{fw82}).

An independent check of the mass loss can be obtained from mid-IR photometry
looking for indications of an infrared excess due to dust emission. NGC\,1846
was covered by the SAGE mid-IR survey (Meixner et al.\,\cite{sage06}). Blum et al.
(\cite{Blum07}) find [8]-[24] as an indicator for mass loss. For five of the carbon
stars in our sample (LE1, LE2, LE3, LE6 and LE11) the corresponding photometry
can be found in the SAGE point source data base. All of them show an [8]-[24] colour
typical for low mass loss stars. We recall that Tanab\'{e} et al. (\cite{T98}, \cite{T04}) did not
find any dust enshrouded objects in this cluster. For two sources Tanab\'{e} et al. (\cite{T04})
list no identification from the literature. Their object 16 is obviously the variable LW2.
Object 8, the weakest object in their list of mid infrared sources, corresponds to LW4.

Using the SAGE list of point sources
we checked for bright objects at 24\,$\mu$m with no or a very weak optical counterpart.
Indeed we found three sources fulfilling this criterion, but their membership to the
cluster is not very likely. In the appendix we give a more detailed description of these
three sources. Thus we can derive from our findings that the AGB stars of NGC 1846
lose their mass in a quite short time at the end of the AGB phase.

The location of LE17 in the P-L-diagram remains somehow a mystery. While the two
carbon stars found at $\log L$/L$_{\odot}$ = 3.57 nicely agree with the expected
location of a TP-AGB star close to its luminosity minimum, it is unlikely
that this explanation also holds for LE17, which is the O-rich star slightly below
in the P-L-diagram (Fig.\,\ref{pl_fig}). If it would be in the minimum of its TP cycle,
it would be in a region exclusively occupied by C-rich stars during its maximum.
A possibility may be that the star is located at the "knee" of the fundamental mode
sequence at its maximum. Alternatively, the derived temperature and luminosity of this star
may be wrong due to higher reddening of this object. Circumstellar reddening probably plays
only a minor effect as there is no clear indication for this from mid-IR photometry (SAGE; Tanab\'{e}
et al.\,\cite{T04}). We further note that the star
nicely fits onto the giant branch of NGC 1846 (Fig.\,\ref{hrd}), thus there is no indication for a
significant reddening. For the same reason we may safely assume that the star is indeed a member
of the cluster. Further investigations
are required to understand the behaviour of this star.

\section{Summary}
 
Twenty two long period variable stars have been detected in the LMC cluster NGC 1846. Most of them
were previously known AGB stars in this cluster. 
We have modelled the pulsation of O-rich and C-rich AGB stars in NGC\,1846,
taking into account the effects of the C/O ratio on the stellar structure.
The stars fit
nicely on several parallel logP-logL-relations. Similar to 47 Tuc (Lebzelter \& Wood
\cite{LW05}), the LPVs populate the first and second overtone sequence at lower luminosities,
while the more luminous stars are found on the first overtone and fundamental sequence.
In addition to the separation by luminosity, there is also a difference related to the
atmospheric chemistry.
The O-rich AGB variables are found to be first and second overtone pulsators.  
We show for the first time that the period of pulsation is increased when an AGB
star turns into a carbon star because of the expansion of the star and the
lowering of $T_{\rm eff}$.  The carbon stars are shown to fall on two sequences
in the P-L-diagram, corresponding to pulsation in the fundamental and first
overtone modes.  Oxygen rich models would give a very poor fit to the PL
relation for C stars and would predict fundamental mode pulsation only.  We
find that there has been no significant mass reduction from mass loss right to
the tip of the observed AGB in NGC\,1846. This is in agreement with mid-IR data.
The mass of the AGB stars is
$\sim$1.8 M$_{\odot}$, corresponding to a cluster age of $1.4\times10^{9}$
years. This is the first determination of the age of a cluster based on the pulsational
properties of its AGB variables and in excellent agreement with the value derived by Mackey \&
Broby Nielsen (\cite{MB07}) from isochrone fitting. However, the separation of the two populations
found by the latter authors is too small to be detectable in the pulsational behaviour of the AGB stars.

\begin{acknowledgements}
TL has been funded by the Austrian FWF under project P18171-N02.
PRW has been partially supported in this work
by the Australian Research Council's Discovery Projects funding scheme (project number DP0663447).
This paper utilizes public domain data obtained by the MACHO Project, jointly funded 
by the US Department of Energy through the University of California, Lawrence Livermore 
National Laboratory under contract No. W-7405-Eng-48, by the National Science 
Foundation through the Center for Particle Astrophysics of the University of California 
under cooperative agreement AST-8809616, and by the Mount Stromlo and Siding Spring 
Observatory, part of the Australian National University. This research is partly based
on data obtained with the 1.3m telescope at CTIO, National Optical Astronomy Observatories,
which is operated by the SMARTS consortium.
\end{acknowledgements}

{}

\begin{appendix}
\section{3 very red sources in the field of NGC 1846}
The SAGE images show three sources in the area of NGC 1846 that are very prominent at 24\,$\mu$m  
and have no clear counterpart in the visual. 
Table \ref{IRsources} gives coordinates and flux values for the three objects. 

\begin{figure}
\centering
\resizebox{\hsize}{!}{\includegraphics{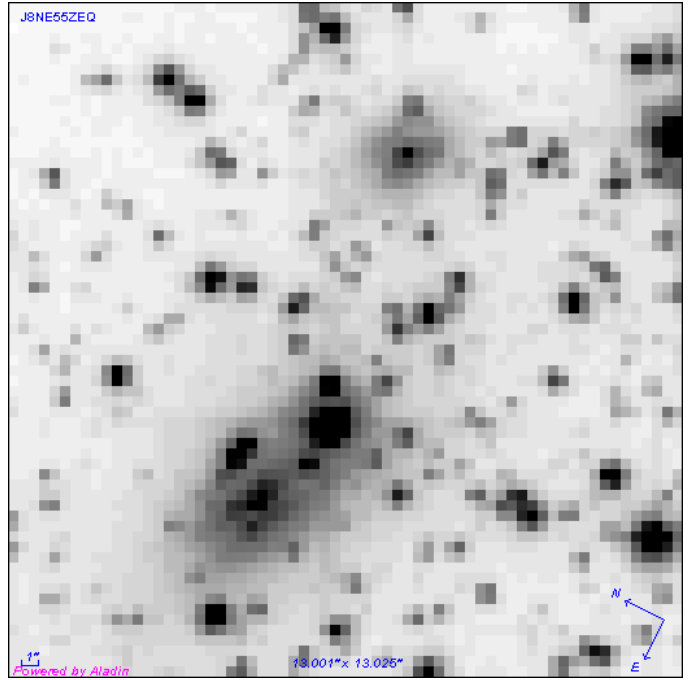}}
\caption{HST/ACS image showing source 1 (upper center) and 2 (lower center), respectively. 
The size of the image is approximately
13 x 13 arcsecs.}
\label{ap12}
\end{figure}

\begin{figure}
\centering
\resizebox{\hsize}{!}{\includegraphics{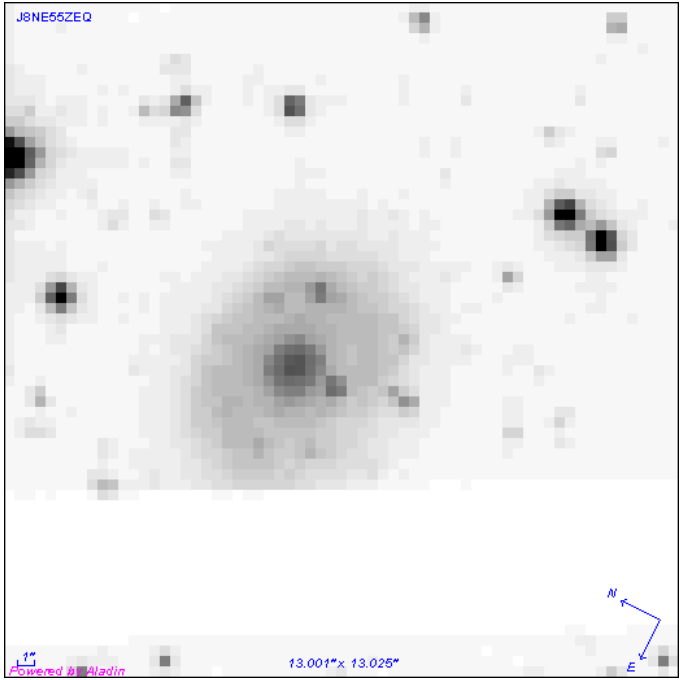}}
\caption{HST/ACS image of source 3 (in the middle above the gap). The size of the image is approximately
13 x 13 arcsecs.}
\label{ap3}
\end{figure}

All three sources are extended as can be seen on the F814W band HST/ACS images available in the
HST data archive (program ID: 9891 , PI: Gerard Gilmore) and shown in Figs.\,\ref{ap12} and \ref{ap3}, respectively.
Source 1 and 2 are located much closer to the cluster center than source 3. 
The location relative to the cluster is also visible from Fig.\,1 of Mackey \& Broby Nielsen (\cite{MB07},
where source 3 is located in the right part of the image just above the gap and source 2 on the upper left part.
None of the three sources could be identified with any object in the list of Tanab\'{e} et al.\,(\cite{T98}, \cite{T04}).

\begin{figure}
\centering
\resizebox{\hsize}{!}{\includegraphics{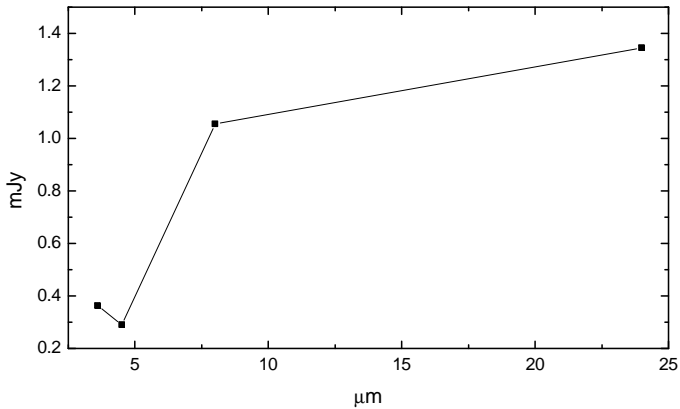}}
\caption{Mid-IR energy distribution of source 1 from SAGE data.}
\label{IRsource1}
\end{figure}

Source 1 has no 
detectable counterpart at 2.2\,$\mu$m but probably at 3.6, 4.5 and 8\,$\mu$m. The energy distribution
is plotted in Fig.\,\ref{IRsource1}. The object is of rather circular shape with an approximate diameter
of 2 arcseconds.

Source 2 looks elongated and is a little bit difficult to separate from a nearby star, probably a RGB
star of NGC 1846 according to its 2MASS $J$ and $K$ values.
It is below the SAGE detection limit at 3.6\,$\mu$m, 4.5\,$\mu$m and
5.8\,$\mu$m, respectively. A source is detected at 8\,$\mu$m at a flux level of 0.585\,mJy. The object
has a size of about 3 to 4 arc seconds.

Source 3 is of slightly elongated shape with an approximate diameter of 5 arc seconds. The image reveals
a central concentration and a few spots possibly related to this source. The central source is not
present in the 2MASS point source catalogue and is also not visible on the 2MASS $K$ frame. This object
is most likely a background galaxy.
Interpretation of sources 1 and 2 is more difficult. We suspect that due to their sizes and redness
both are background galaxies. They seem to be far too weak at wavelengths shortward of 20\,$\mu$m
to represent dust enshrouded objects within the cluster.

\begin{table}
\caption{Infrared sources in the field of NGC 1846. Coordinates and fluxes are taken from the SAGE catalogue.}
\label{IRsources}
\centering
\begin{tabular}{l c c c}
\hline\hline
Ident & RA (2000) & DE (2000) & F$_{24\mu}$ (mJy)\\
\hline
Source 1 & 05:07:33.21 & -67:27:06.2 & 1.346\\
Source 2 & 05:07:34.42 & -67:27:07.2 & 0.799\\
Source 3 & 05:07:34.94 & -67:29:01.8 & 0.875\\
\hline
\end{tabular}
\end{table}

\end{appendix}

\end{document}